\documentclass[5p]{elsarticle}
\pdfoutput=1

\usepackage[parfill]{parskip} 

\usepackage[utf8]{inputenc} 
\usepackage{geometry} 
\usepackage{graphicx} 
\usepackage{url}
\usepackage[colorlinks=true,linkcolor=blue,citecolor=blue]{hyperref}%
\usepackage{booktabs} 
\usepackage{array} 
\usepackage{paralist} 
\usepackage{verbatim} 
\usepackage{caption}
\usepackage{subcaption} 
\usepackage{epstopdf}
\usepackage{placeins}
\usepackage{fancyhdr} 
\usepackage{amsmath}
\usepackage{amssymb}
\usepackage{enumerate}
\usepackage[activate={true,nocompatibility},final,tracking=true,kerning=true,spacing=true,factor=1100,stretch=10,shrink=10]{microtype}
\usepackage{color}
\usepackage{xcolor}
\usepackage[expproduct=tighttimes]{siunitx} 
\usepackage{bbm} 
\usepackage{upgreek}
\usepackage{relsize} 

\usepackage{ctable}\newcommand{\thickmidrule}{\midrule[\heavyrulewidth]} 

\renewcommand{\b}[1]{{\boldsymbol{#1}}} 

\newcommand{\pd}[2]{\frac{\partial #1}{\partial #2}}

\newcommand{\fbm}{\ensuremath{{f_\text{bm}}}}
\newcommand{\fvas}{\ensuremath{{f_\text{vas}}}}

\newcommand{\bm}{\text{bm}}
\newcommand{\vas}{\text{vas}}
\newcommand{\rve}{\text{\textsmaller{ROI}}}

\newcommand{\bv}{\text{\textsmaller{BV}}}
\newcommand{\bs}{\text{\textsmaller{BS}}}
\newcommand{\tv}{\text{\textsmaller{TV}}}

\newcommand{\micro}{\text{micro}}
\newcommand{\cmicro}{\ensuremath{\mathbbm{c}^\text{micro}}}

\newcommand{\sv}{\ensuremath{S_{\textrm{V}}}}
\newcommand{\ob}{\text{\textsmaller{O}b}}
\newcommand{\oc}{\text{\textsmaller{O}c}}

\newcommand{\kform}{\text{$k_\text{form}$}} 
\newcommand{\kres}{\text{$k_\text{res}$}}

\newcommand{\SEDcell}{\text{$\Psi_0$}}
\newcommand{\SEDcellmicro}{\text{$\Psi_0^{\textrm{cell}}$}}

\newcommand{\bstrut}[1]{\rule[-#1]{0pt}{#1}} 
\newcommand{\f}{\ensuremath{f}}

\journal{arXiv}

\begin{document}

\begin{frontmatter}

\title{Towards a cell-based mechanostat theory of bone: the need to account for osteocyte desensitisation and osteocyte replacement}

\author[mon]{Chlo\'e Lerebours}
\author[mon]{Pascal R. Buenzli}

\address[mon]{School of Mathematical Sciences, Monash University, Clayton VIC 3800, Australia.}

\begin{abstract}
Bone's mechanostat theory describes the adaptation of bone tissues to their mechanical environment. Many experiments have investigated and observed such structural adaptation. However, there is still much uncertainty about how to define the reference mechanical state at which bone structure is adapted and stable. Clinical and experimental observations show that this reference state varies both in space and in time, over a wide range of timescales. We propose an osteocyte-based mechanostat theory that links various timescales of structural adaptation with various dynamic features of the osteocyte network in bone. This theory assumes that osteocytes are formed adapted to their current local mechanical environment through modulation of morphological and genotypic osteocyte properties involved in mechanical sensitivity. We distinguish two main types of physiological responses by which osteocytes subsequently modify the reference mechanical state. One is the replacement of osteocytes during bone remodelling, which occurs over the long timescales of bone turnover. The other is cell desensitisation responses, which occur more rapidly and reversibly during an osteocyte's lifetime. The novelty of this theory is to propose that long-lasting morphological and genotypic osteocyte properties provide a material basis for a long-term mechanical memory of bone that is gradually reset by bone remodelling. We test this theory by simulating long-term mechanical disuse (modelling spinal cord injury), and short-term mechanical loadings (modelling daily exercises) with a mathematical model. The consideration of osteocyte desensitisation and of osteocyte replacement by remodelling is able to capture the different phenomena and timescales observed during the mechanical adaptation of bone tissues, lending support to this theory.

\end{abstract}


\end{frontmatter}

\section{Introduction}
The mechanostat theory of bone proposed by Frost \cite{Frost1987, Frost2003} describes the adaptation of bone tissues to their mechanical environment by a simple feedback loop. Regions of bone experiencing high mechanical loads become consolidated, while regions of bone experiencing low mechanical loads are removed. Many clinical situations and experiments have exhibited such bone adaptations over a wide range of timescales, from short-term bone gain following daily exercises, to long-term bone loss following spinal cord injury~\citep{Vico2000, Robling2001, Ehrlich2002, Burr2002, Turner2004, Eser2004, Thompson2012}. Several computer algorithms have been developed, in which bone density or microstructure adapts in response to mechanical loads~\citep{Frost1987, Beaupre1990, Huiskes2000, Ruimerman2005, Garcia-Aznar2005, Lerebours2015}. The mechanostat theory is of great importance in biomechanics studies that aim to understand the role of mechanics in the deterioration of bone tissues and the influence of physical activity for the preservation of bone with age \citep{Goel1995, Lerebours2015}. Long-term mechanical adaptation is particularly significant to implant integration and stability~\citep{VanRietbergen1993, Turner2005}, as well as scheduling of brace displacement, such as orthodontic braces~\cite{Chou2008}.

A mechanostat relies on the definition of a mechanical reference state (a setpoint, or broader ``lazy zone'') above which the tissue experiences mechanical overuse, and under which the tissue experiences mechanical disuse. Complex spatial and temporal dependences of bone adaptation imply that bone's mechanical setpoint varies both in space and in time~\cite{Turner1999a, Skerry2006}. For example, bone tissue near the neutral axis of long bones is mostly unloaded, but is not resorbed. Also, load timing and rest periods influence bone adaptation, and lead to load-history-dependent bone structures \citep{Forwood1995, Schriefer2005}.

Osteocytes are cells embedded in bone matrix during bone formation. These cells play a fundamental role in mechano\-sensation and mechano\-transduction in bone \citep{Turner1994, Marotti2000, KnotheTate2003, Adachi2010, Bonewald2011, Buenzli2015}. However, no mechanostat theory has yet captured the multiple time\-scales of bone adaptation while accounting for the cellular basis of mechanosensation. While most computer models assume a fixed, universal setpoint, some have considered setpoints that relax to current mechanical stimulus by cell accommodation, e.g. through cytoskeleton reorganisation or receptor desensitisation. Turner \cite{Turner1999a} proposed to redefine the setpoint dynamically according to the duration and strength of the mechanical stimulus experienced, to represent desensitisation of osteocytes. This theory leads to a time-dependent and space-dependent reference mechanical state. It has been implemented in a number of bone remodelling models~\citep{Schriefer2005,Garcia-Aznar2005}, which exhibit loading-history-dependent results that agree with experimental observations~\citep{Schriefer2005}. However, in these models, cells re-sensitise within 25 to 500 days. These times are not representative of the biological process of cell accommodation, which occurs within hours~\citep{Ehrlich2002,Adachi2009,Burra2010,Bergmann2011}. There are other physiological timescales in the adaptation of bone that could explain such slower responses.

In this paper, we argue that a mechanostat theory of bone must account for three different physiological responses. (i) Rapid, reversible osteocyte responses ($<$24\,hours) due to a mismatch between setpoint and actual mechanical state. This includes both osteocyte signalling to bone-forming cells (osteoblasts) and bone-resorbing cells (osteoclasts) and the desensitisation of osteocytes to the mechanical stimulus by a rapid modulation of mechanical sensitivity. Such rapid responses may occur by changes in protein expression, receptor desensitisation, and reorganisation of the actin cytoskeleton and of dendritic cell processes~\citep{Adachi2009,Burra2010,Bergmann2011,Rubin1999,Dallas2010, Turner2004}. (ii) The adaptation of bone structure by bone formation and bone resorption (weeks--years). This corresponds to the response of the conventional mechanostat. (iii) The replacement of osteocytes during bone remodelling, which enables a slower adaptation of mechanical sensitivity (weeks--years). A long-term mechanical memory of bone may be materialised in long-lasting morphological and genotypic osteocyte properties. This long-term memory is replaced during remodelling, at a rate commensurate with bone turnover rate, which depends in particular on bone anatomical regions.

These considerations define an osteocyte-based reference mechanical state that is inhomogeneous and dynamic over several timescales. We test this theory by simulating long-term mechanical disuse and exercise regimens with a mathematical model of bone remodelling. By accounting for the replacement of mechanical memories by bone remodelling, our proposed theory includes an additional degree of freedom of adaptation able to resolve spatial and temporal shortcomings of previous mechanostat theories.

\section{Osteocyte-based mechanostat theory}\label{section_proposal_mechanostat}
\begin{figure*}
\begin{center}
\includegraphics[trim = 0cm 4.8cm 0.3cm 1.2cm, clip=true, width=0.9\textwidth]{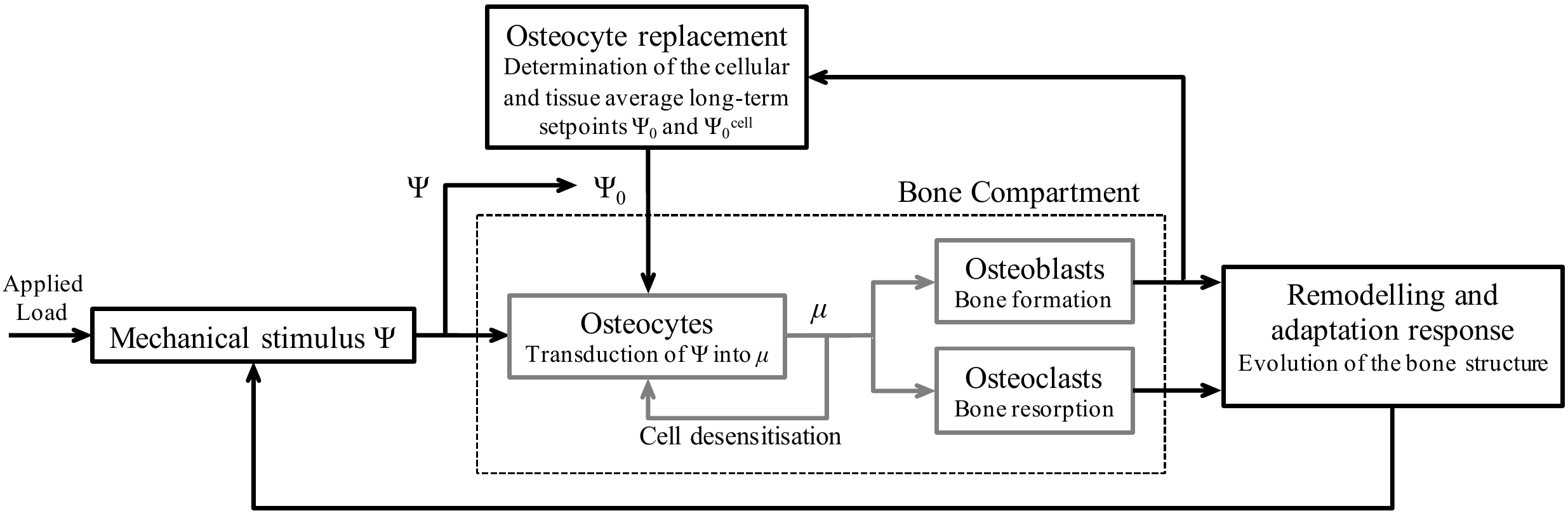} 
\caption{Conceptual diagram of the proposed osteocyte-based mechanostat theory, which includes: (i) a site-specific mechanical stimulus $\Psi$ (e.g., the strain energy density) determined by the load distribution in the bone structure; (ii) the transduction of $\Psi$ into a biochemical stimulus $\mu$ by the osteocytes; (iii) the bone formation and bone resorption processes induced in responses to $\mu$; (iv) the desensitisation of the osteocytes in response to $\mu$; and (v) the determination of the cellular and tissue average long-term setpoints $\SEDcell$ and $\SEDcellmicro$ during new bone formation.}
\label{model_diagram}
\end{center}
\end{figure*}

Bone loss due to mechanical disuse is observed in human and animal studies after prolonged bed rest, spaceflight missions, and spinal cord injuries \citep{Collet1997,Vico2000, Eser2004, Zehnder2004, Gross2010, Jaworski1986, Jee1990}. Bone loss is significant after only a few weeks of mechanical disuse, but it can continue for several years. Spinal cord injury patients lose bone during at least the first 8 years post injury~\citep{Eser2004, Zehnder2004}. In contrast, bone gain due to mechanical overuse occurs only if several conditions are met. Bone gain depends on the frequency, strain rate, amplitude, duration of the loading, and the interpolation of rest periods~\citep{Sibonga2007,Ehrlich2002, Srinivasan2007, Thompson2012,Turner1998}. Bone gain is enhanced by rest periods of (i) 14~seconds, due to the viscoelastic recovery of the tissue~\citep{Turner1998,Robling2001, Turner2004, Srinivasan2007}; (ii) 8~hours, due to the re-sensitisation of the mechanosensing cells to previous levels of mechanical loading~\citep{Turner1998,Robling2001, Burr2002, Turner2004}; and (iii) 5~weeks, attributed to learning and memory circuits in the nervous system~\citep{Turner1998,Turner2004, Saxon2005}.

A comprehensive mechanostat theory of bone needs to incorporate these various physiological processes and related timescales. We will show in this paper that osteocyte desensitisation and osteocyte replacement enable to account both for short-term responses (hours) and for mid-to-long-term responses (weeks to years).

\subsection{Osteocyte desensitisation (short-term response)} \label{subsection2.1_cell_desensit}
Mechanically stimulated osteocytes desensitise their response to stimulus within a few hours~\citep{Robling2001, Turner2004, Burr2002}. This may occur by creation/removal of tethering elements attaching the cell to the bone surface (see Fig.~\ref{cell_desensitisation_drawing}), by rearrangement of the cytoskeleton's actin network, by intracellular mechanisms such as cell surface receptor desensitisation, or by rearrangment of dendritic cell processes and osteocyte connections~\citep{Adachi2009,Burra2010,Bergmann2011, Rubin1999,Turner2002, Turner2004, Thompson2012, Dallas2010}. This rapid cellular accommodation corresponds to a short-term modulation of the setpoint. It occurs during an osteocyte's lifetime and is reversible. Clearly, such accommodation cannot be total, otherwise no long-term mechanical adaptation of bone would occur. We hypothesise that osteocytes undergo rapid, but \emph{partial} desensitisation to the mechanical stimulus, after which cell response still occurs, but with reduced intensity. The possibility to re-sensitise osteocytes after 8\,hour rest periods implies that a longer-term memory of a mechanical reference state exists in osteocytes.

\subsection{Osteocyte replacement (mid-term and long-term responses)}
The evolution of bone structure over longer timescales redistributes the mechanical loads carried by the different bone tissues. The mechanical stimulus $\Psi$ sensed by the osteocytes feeds back into the transduction mechanisms that govern bone changes. The slow co-evolution of bone structure and mechanical stimulus is precisely what Frost's mechanostat theory describes. We argue that at this long timescale, \emph{no adaptation of bone consistent with all the experimental observations can occur by changes in bone structure only. A change in time of the reference mechanical state itself is required at long timescales}. Indeed, past the initial transient desensitisation, an adapted bone structure is reached only when the current mechanical state $\Psi$ equals the long-term mechanical setpoint $\SEDcell$ everywhere. If $\SEDcell$ is fixed in time but $\Psi$ is varied, bone structure will evolve by formation and resorption so as to bring $\Psi$ towards $\SEDcell$. If the neutral axis moves (e.g. due to load redistribution during age-related bone loss), this will always lead to total loss of bone at the new neutral axis, where $\Psi=0$. These limitations of Frost's mechanostat have been raised before to emphasise the need to account for the spatial and temporal dependences of the mechanical reference state itself~\cite{Turner1999a, Skerry2006}.

Because setpoint modulation by cell accommodation occurs over short timescales, we are led to consider other changes that occur during the evolution of bone tissues, namely, \emph{slow changes in bone material properties}~\citep{Buenzli2015a,Buenzli2015b}. Since osteocytes are embedded within bone matrix during bone formation, the long-term mechanical setpoint $\SEDcell$ may be viewed as an osteocyte-specific tissue property. We propose that $\SEDcell$ is encoded by long-lasting morphological and genotypic properties of osteocytes. We hypothesise that osteocytes are formed `adapted' to their local mechanical environment in such a way that $\SEDcell = \Psi$ at the time of osteocyte formation, and that this long-term memory $\SEDcell$ of the mechanical state $\Psi$ persists until the osteocyte's removal. While these hypotheses remain to be validated experimentally, they are supported by the observation that the osteocyte and lacuno-canalicular networks (the micro-pores within which osteocytes live) have different morphologies in different mechanical environments~\citep{Mullender1996, VanOers2015}. Microscopic stress concentration effects around the lacuna and canaliculi are likely to affect sensation by the osteocyte of the surrounding mechanical stimulus $\Psi$~\citep{Burger1999, Dong2014,VanOers2015,Vatsa2008,vanHove2009, Steck2005}, and therefore to modulate a long-term component of the setpoint.

\subsection{Osteocyte-based mechanostat}
The above considerations lead us to conceptualise an osteocyte-based mechanostat as follows (see Figure~\ref{model_diagram}):
\begin{enumerate}
    \item The distribution of applied muscle and tendon forces in the bone structure leads to a site-specific \emph{mechanical stimulus} $\Psi$ sensed by the osteocytes;
    \item Osteocytes transduce the mechanical stimulus $\Psi$ into a \emph{biochemical stimulus} $\mu$ signalling overload if \mbox{$\mu>0$} and underload if \mbox{$\mu<0$};
    \item The biochemical stimulus $\mu$ feeds back to osteocytes, resulting in their partial desensitisation;\label{item:desensitisation}
    \item The biochemical stimulus $\mu$ stimulates bone formation in overload and bone resorption in underload;
    \item Formation and resorption modify bone structure, leading to a redistribution of the applied forces in the tissue;
    \item Bone formation during bone renewal replaces tissue material properties, including osteocytes. Osteocyte replacement resets the long-term reference state $\SEDcell$ to the current mechanical stimulus $\Psi$. After osteocyte desensitisation transients, overload is equivalent to $\Psi>\SEDcell$ and underload is equivalent to $\Psi<\SEDcell$.
\end{enumerate}

In summary, the long-term setpoint $\SEDcell$ determined during the formation of an osteocyte is a long-lasting memory of the present mechanical stimulus, that gives a cellular basis to a site-specific mechanical reference state. This memory is gradually lost over long timescales, and replaced by newer mechanical memories, by the slow process of bone remodelling which renews tissue material properties and replaces osteocytes. 

\section{Mathematical model}\label{Sec_math_model}
To test the implications of this mechanostat theory, we implement it in a generic mathematical model of bone remodelling. The model is briefly summarised here. See~\ref{sec:model-description} for a detailed presentation as well as a discussion of parameters and their calibration.

We consider a region of interest (\rve) of the tissue experiencing an internal compressive force $F(t)$. The bone volume fraction $\fbm$ of the \rve\ evolves according to the densities of active osteoblasts~$\ob$ and active osteoclasts~$\oc$:
\begin{align}\label{ODE_fbm}
\frac{\partial}{\partial t} \fbm (t) = \kform ~\ob - \kres ~\oc,
\end{align}
where $\kform$ and $\kres$ are the cell secretory and dissolution rates~\citep{Martin1972,Pivonka2008,Buenzli2013}. Biochemical, geometrical, and mechanical regulations are assumed to influence the populations of active osteoblasts and osteoclasts, see Eqs~\eqref{kfOB}--\eqref{krOC}. Mechanical regulation is modelled by terms that depend on the osteocyte population and on the transduced biochemical stimulus $\mu$.  In a healthy steady state at mechanical equilibrium, bone volume fraction is constant, but continually turned over with rate $\kform\ob = \kres\oc$~\citep{Lerebours2015}.

\begin{figure}
\begin{center}
\begin{subfigure}{0.32\textwidth}
\includegraphics[width=\textwidth]{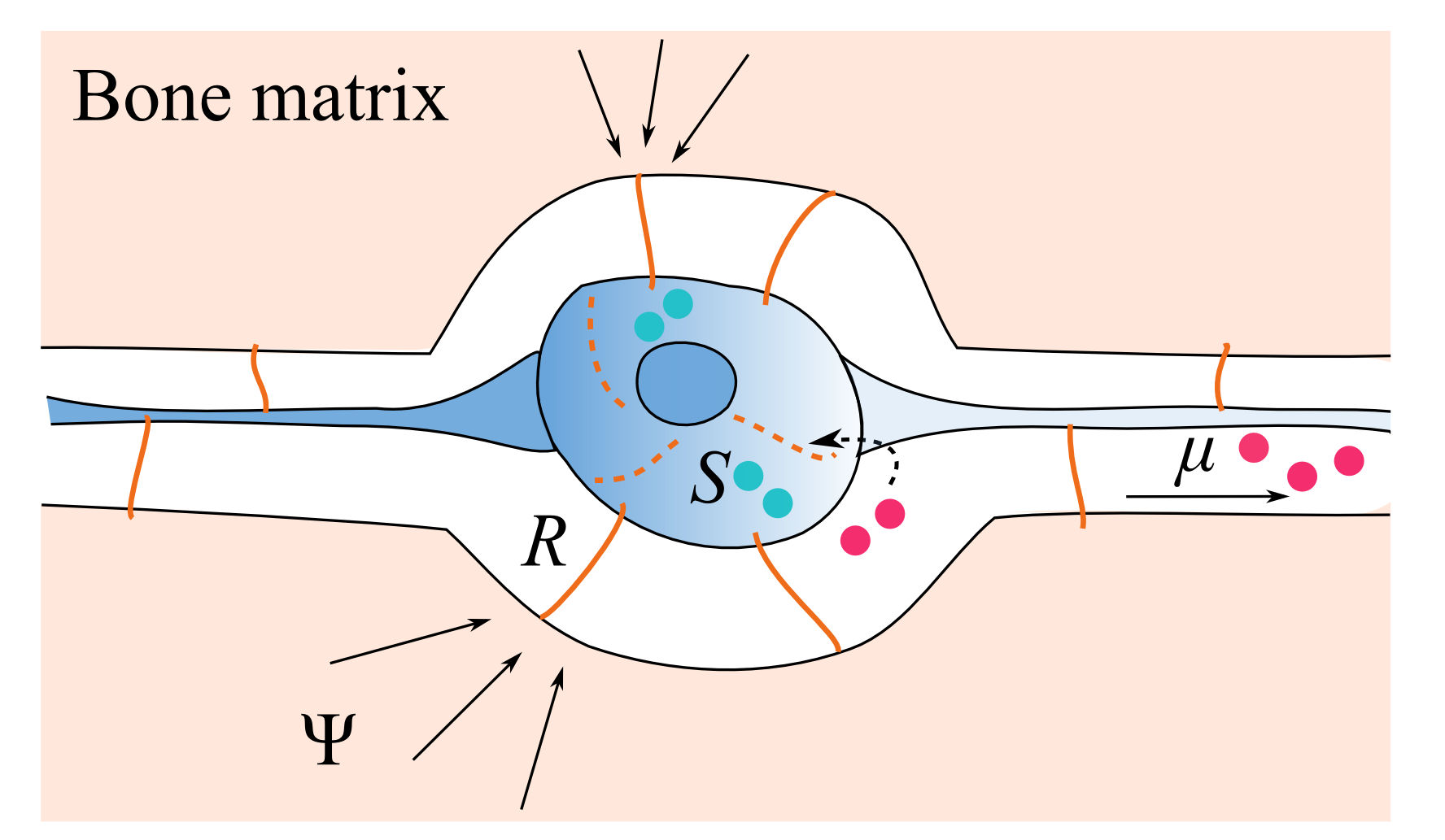}
\caption{}
\label{cell_desensitisation_drawing}
\end{subfigure}
\begin{subfigure}{0.14\textwidth}
\includegraphics[trim = 2.4cm 19cm 10cm 0cm, clip=true, width=\textwidth]{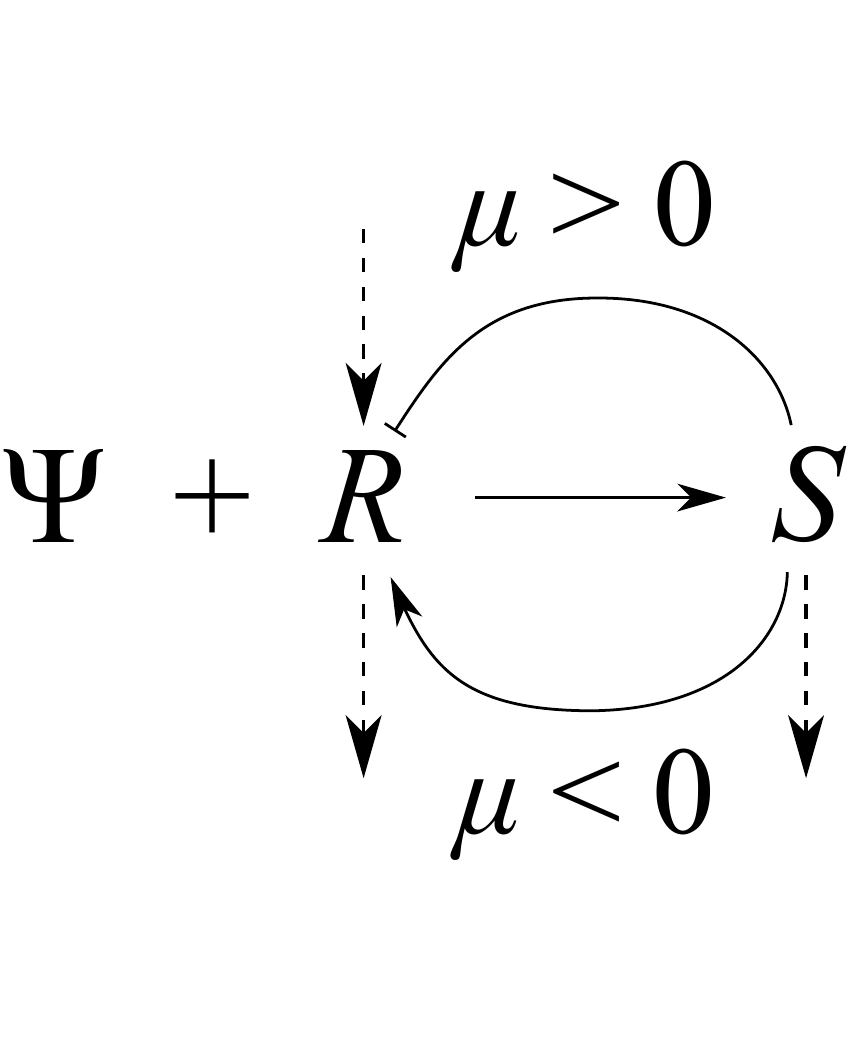}
\caption{}
\label{cell_desensitisation_equation}
\end{subfigure}
\caption{(a) Mechanotransduction and cell desensitisation model. An osteocyte is tethered in a lacuna by elements $R$ that allow the mechanical stimulus $\Psi$ to be sensed and transduced into a biochemical compound $S$ and a  transduced biochemical stimulus $\mu$. This stimulus feeds back into the osteocyte to create or remove tethering elements $R$, and signals osteoclasts and osteoblasts through the canalicular network. (b) Receptor--ligand reaction describing the model.}
\label{cell_desensitisation_model}
\end{center}
\end{figure}

Mechanobiological transduction and osteocyte desensitisation are described by a cellular scale model of receptor--ligand signalling and trafficking summarised in Figure~\ref{cell_desensitisation_model}. The mechanical stimulus $\Psi$ is assumed to be transduced by mechano-receptors $R$ into an intracellular agent $S$. An intracellular cascade then generates an extracellular biochemical stimulus $\mu$ such that if $S>S_0$, $\mu>0$ and if $S<S_0$, $\mu<0$, where $S_0$ is a reference number of molecules~$S$ involved in the definition of the long-term reference state $\SEDcell$, see Eq.~\eqref{sedcell-def}. Cell accommodation to $\Psi$ is modelled by a feedback of $\mu$ onto the number of mechano-receptors~$R$, or equivalently, onto the mechano-receptors' sensitivity. 

The mechanostat's long-term reference state $\SEDcell$ is defined by osteocyte-specific morphological and genotypic parameters. These parameters are set such that a new osteocyte is mechanically adapted to the stimulus $\Psi$ that prevails during its formation, i.e., such that $\SEDcell=\Psi$. To evaluate how the replacement of osteocytes influences the average value of $\SEDcell$ in the \rve, we used microscopic equations governing the evolution of tissue properties during modelling and remodelling, and averaged them in the \rve\ using a mean-field approximation~\citep{Buenzli2015b}, giving:
\begin{align}
\frac{\partial}{\partial t} \SEDcell &= \frac{1}{\fbm} ~\kform \ob ~(\Psi - \SEDcell). \label{SEDcell_ODE}
\end{align}
This equation describes the gradual replacement of $\SEDcell$ by the current value of the mechanical stimulus $\Psi$ at a rate proportional to the bone formation rate $\kform \ob$. This can be seen as keeping a memory of the current mechanical stimulus into newly formed bone~\citep{Buenzli2015a,Buenzli2015b}. The prefactor $\frac{1}{\fbm}$ means that it is quicker to average out the pre-existing value of $\SEDcell$ by the newer value $\Psi$ when there is little bone. 

Finally, the mechanical stimulus $\Psi$ is assumed to be the strain energy density. It is determined from the applied force $F(t)$ and bone volume fraction $\fbm$, see~\ref{subsection_meca}.

We note here that \SEDcell\ is a local average over the \rve\ of individual osteocytic values $\SEDcellmicro$ (see~\ref{subsection_cell_desensit}). This averaging represents the integrative capacity of the osteocyte network. It also ensures the stability of the model even when nearby osteocytes send contradictory signals. This is a well-known issue that can lead to ``checkerboard" instabilities in the absence of averaging~\citep{Weinans1992, Mullender1994}.

\section{Results}\label{sec_results}
The equations governing the evolution of $\fbm$, $\mu$, $\Psi$, and $\SEDcell$ were simulated numerically under loading scenarios corresponding to mechanical disuse and exercise regimens.

\subsection{Mechanical disuse}
\begin{figure}[t!]
\begin{center}
\includegraphics[trim = 0.5cm 0.3cm 0.4cm 0.9cm, clip=true,width=0.5\textwidth]{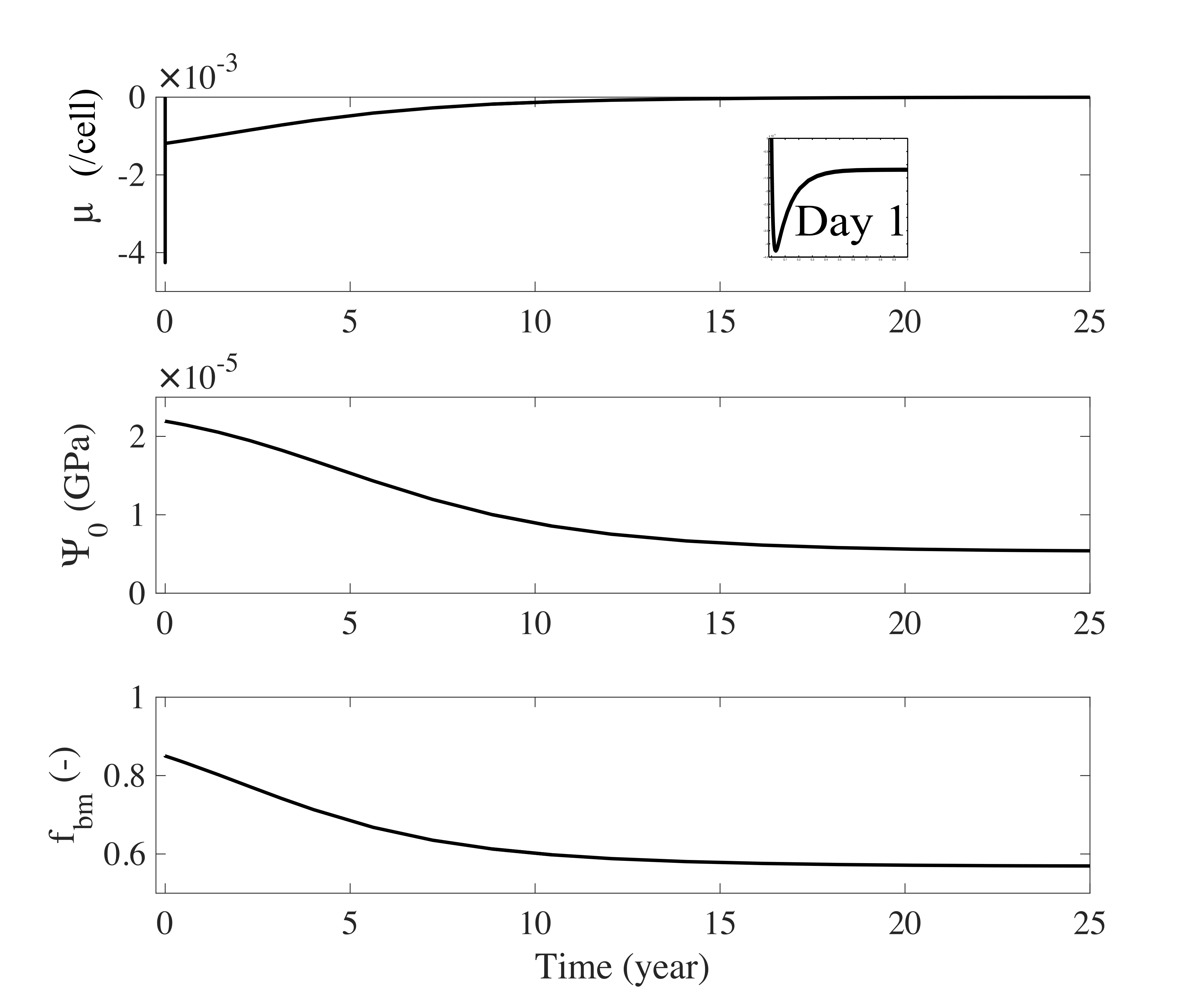}
\caption{Simulation of 25 years of mechanical disuse. Time evolution of (a) the transduced biochemical stimulus $\mu$, with a close-up view of the first day of disuse (inset); (b) the long-term setpoint \SEDcell; and (c) the bone volume fraction, \fbm.}
\label{Disuse25y}
\end{center}
\end{figure}

Figure \ref{Disuse25y} depicts the time evolution of the transduced biochemical stimulus $\mu$, the long-term setpoint \SEDcell, and the bone volume fraction \fbm\ when simulating 25 years of mechanical disuse modelled by a reduction in initial force $F$ to $F_{\textrm{disuse}} = F/3$. This situation is representative of leg paralysis after spinal cord injury. The inset shows that within one day after the onset of mechanical disuse, the biochemical stimulus $\mu$ converges to the quasi-steady state value $\overline\mu$ after a short transient due to the osteocytes' partial desensitisation (see \ref{subsection_cell_desensit}). This quasi-steady state value $\overline\mu$ gradually reduces its intensity over the years, and relaxes to zero after about 10 years. This timescale corresponds to the time required for $\SEDcell$ and $\fbm$ to reach new stationary values, after a total bone loss of 30\%.

\begin{figure}[t!]
\begin{center}
\includegraphics[trim = 0.8cm 1.1cm 0.5cm 0.5cm, clip=true,width=0.5\textwidth]{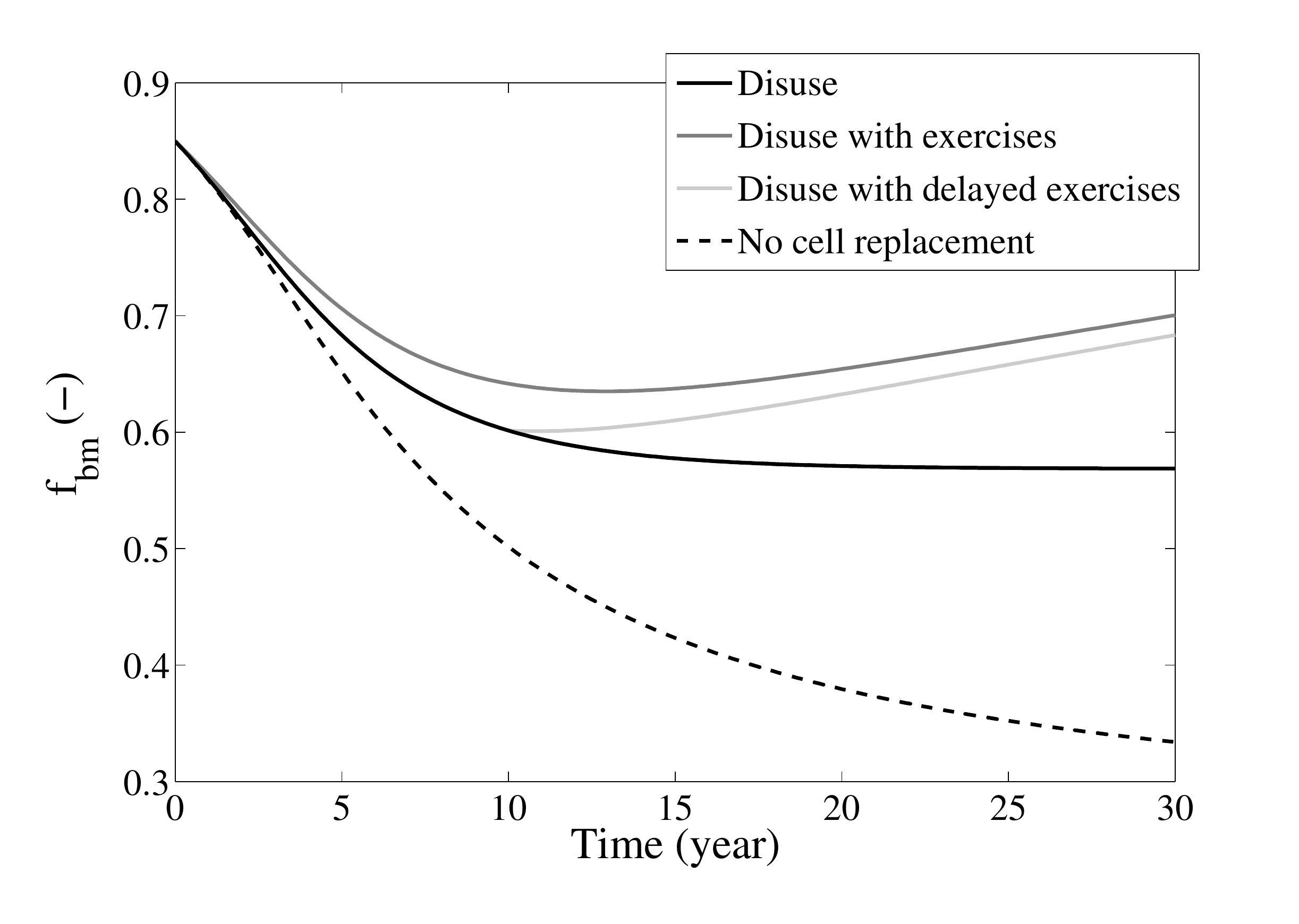} 
\caption{Time evolution of $\fbm$ under mechanical disuse without exercises (solid black), with exercises (solid dark gray), and with delayed exercises starting 10 years after the onset of disuse (solid light gray). The interrupted line correspond to simulations performed without including the replacement of osteocytes.}
\label{Disuse30y_comparison_models}
\end{center}
\end{figure}

Figure \ref{Disuse30y_comparison_models} shows the influence of osteocyte replacement on the relaxation to a new adapted bone microstructure. The solid black line is the same as in Figure~\ref{Disuse25y}, while the interrupted line is obtained by enforcing $\SEDcell(t) \equiv \SEDcell(0)$ at all times, which corresponds to not accounting for the loss of mechanical memory induced by cell replacement. This leads to higher bone loss and no stabilisation of the bone volume fraction within 30 years. Figure~\ref{Disuse30y_comparison_models} also shows the influence of exercise regimens superimposed to the disuse, simulated as daily oscillations in the external force $F(t)$. The force was assumed to increase to the value $1.2 F_\text{disuse}$ during exercise (see also Section~\ref{sec:exercise-regimes}). An exercise regimen does not significantly change the loss of bone in the first 10 years. However, it enables bone to be recovered subsequently. Starting exercises only after 10 years of bone loss does not reduce much the amount of bone recovered.

\subsection{Exercise regimens}\label{sec:exercise-regimes}
\begin{figure}[t!]
\begin{center}
\includegraphics[trim = 0cm 0cm 0.7cm 0.5cm, clip=true,width=0.5\textwidth]{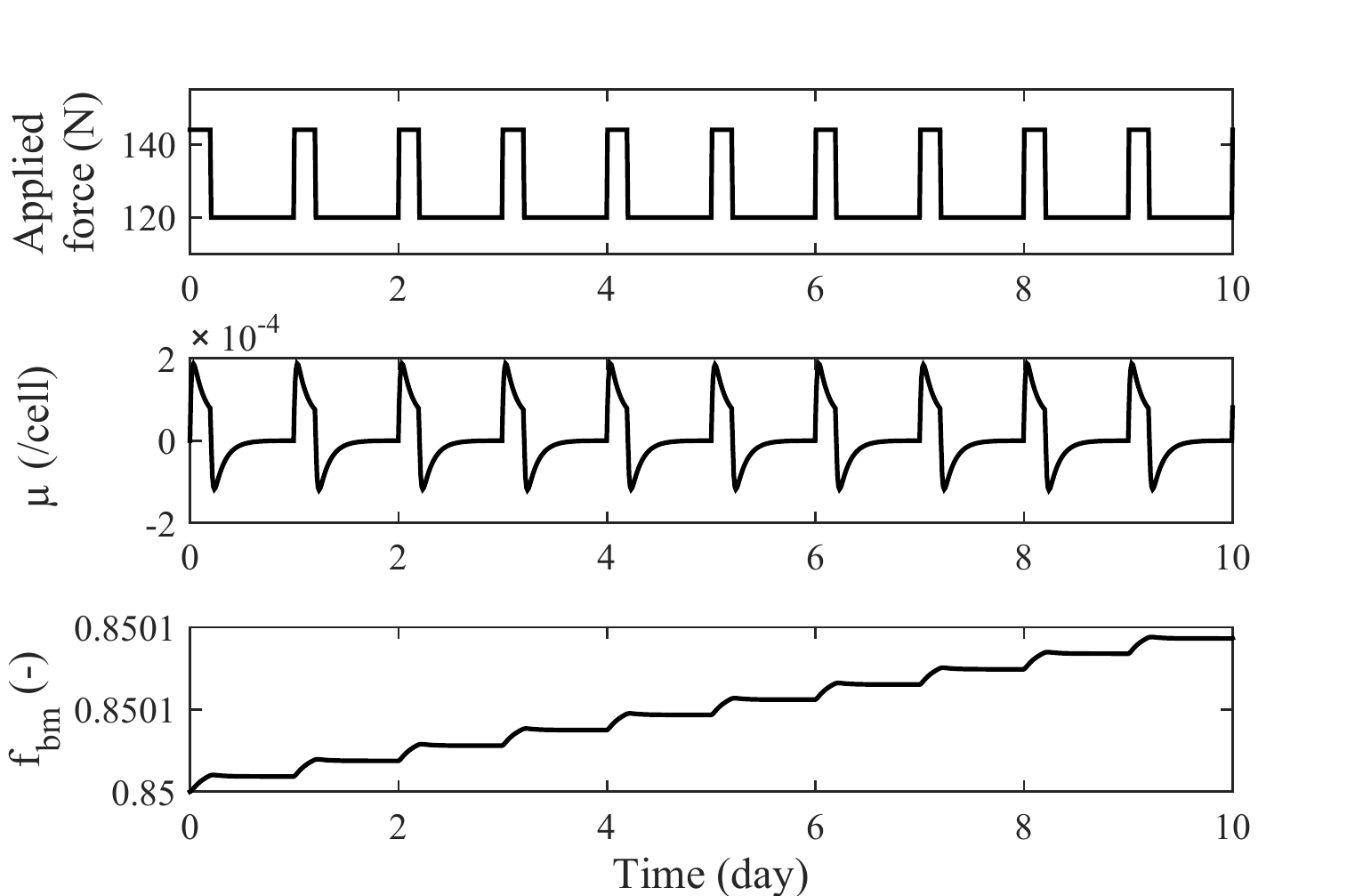} 
\caption{Simulation of 10 days of daily exercise. Time evolution of (a) the applied force; (b) the transduced biochemical stimulus $\mu$; and (c) the bone volume fraction, \fbm.}
\label{Bumping10d}
\end{center}
\end{figure}

We investigated the short-term influence of the number and duration of exercises in an exercise regimen representing an increase in physical activity of a healthy individual. The duration of exercises $T_\text{exercise}$, and the duration of rest periods between exercises $T_\text{rest}$ were varied. Figure \ref{Bumping10d} shows the time evolutions obtained with $T_\text{exercise}=5\,\text{h}$, and $T_\text{rest}=19\,\text{h}$. Each relative increase or decrease in $F(t)$ induces a sudden increase or decrease in $\mu(t)$, respectively, after which $\mu(t)$ relaxes towards a smaller value due to the osteocytes' partial desensitisation. The response of $\mu(t)$ to a relative increase or decrease in $F(t)$ is not symmetrical with respect to the $\mu=0$ axis. These periodic micro-loadings lead to a gradual increase in the bone volume fraction. 

\begin{figure}[t!]
\begin{center}
\includegraphics[trim = 0.1cm 0cm 0.5cm 0.5cm, clip=true,width=0.5\textwidth]{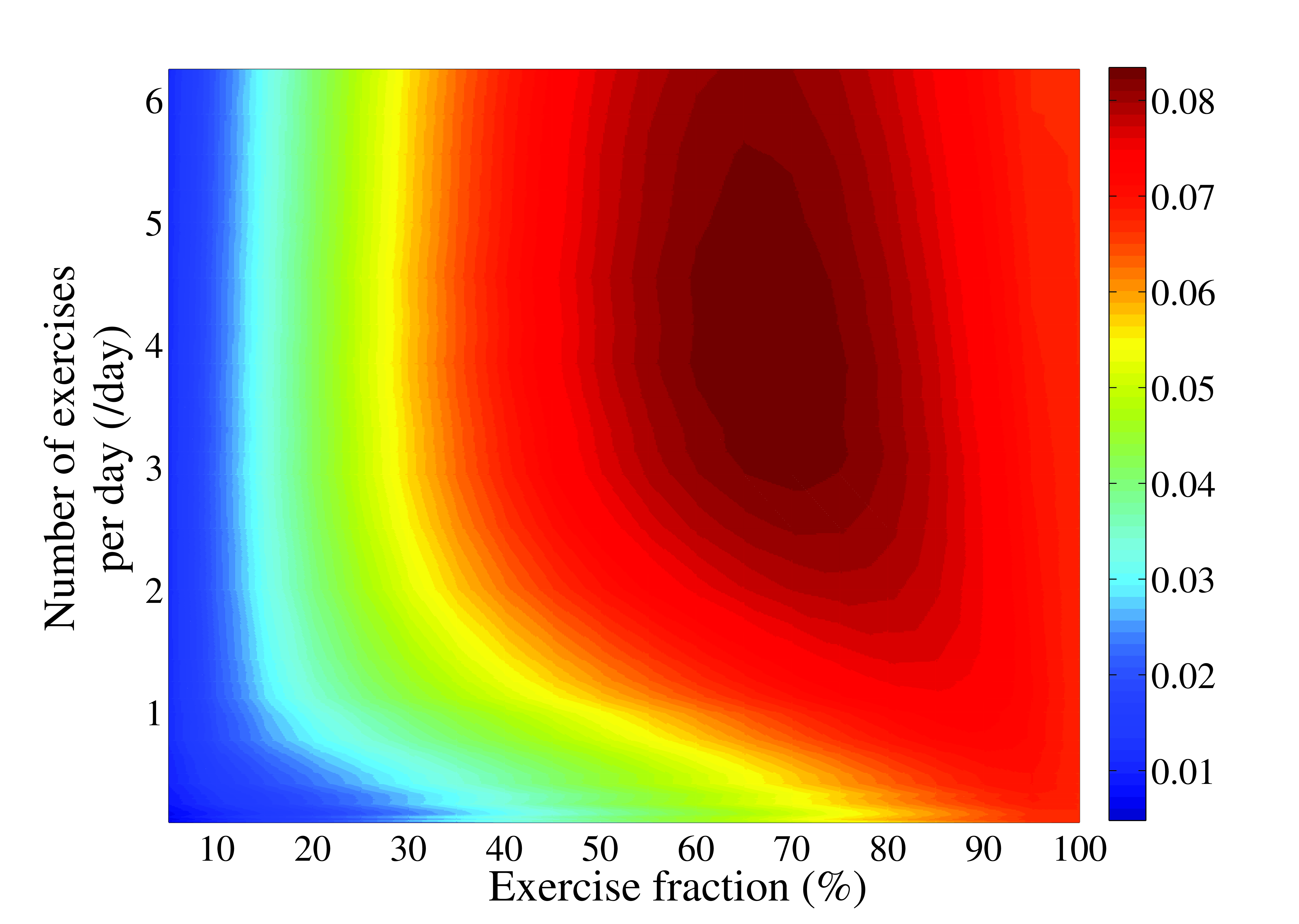}
\caption{Percentage of bone gain after 30 days of exercises depending on the number of exercises per day and the exercise fraction. Note that along a vertical line, the total time exercising is constant.}
\label{Influence_periods}
\end{center}
\end{figure}

Figure \ref{Influence_periods} sums up a parametric study performed by varying $T_\text{exercise}$ and $T_\text{rest}$. The percentage of bone gain after 30 days of exercises is plotted versus the exercise fraction $\frac{T_\text{exercise}}{T_\text{exercise}+T_\text{rest}}$, and the daily number of exercises $\frac{24\,\text{h}}{T_\text{exercise}+T_\text{rest}}$. When exercising 3 times a day with an exercise fraction of 20\%, more bone can be gained by increasing the exercise fraction, i.e., by increasing the total amount of time exercising. However, when exercising once a day with an exercise fraction of 50\%, more bone can be gained by increasing the number of daily exercises, i.e., doing shorter exercises more often. Exercising all day round (100\% exercise fraction) is not optimal, and just as efficient as exercising three times a day with an exercise fraction of 35\%, corresponding to three bouts of exercises of 2.8 hours each.

\section{Discussion}\label{sec_discussion}
A comprehensive mechanostat theory of bone needs to be consistent for every type of bone tissue, and to account for asymmetries in bone responses in mechanical disuse and in mechanical overuse. The two major difficulties in formulating such a comprehensive theory lie in (i) the absence of a one-to-one correspondence between local mechanical environment and microarchitecture, highlighting the dependence on load history; and (ii) the various timescales at which mechanical adaptation occurs, in particular:
\begin{enumerate}
    \item the rapid decrease in bone gain in the absence of rest periods (hours timescale)
    \item the possibility that bone gain may be sustained over longer timescales (days to weeks timescales)
    \item the possibility that bone loss may be sustained over very long timescales (years to decade timescales)
\end{enumerate}

The mechanostat theory of bone we propose is novel on several points: (i) it provides an explicit cellular representation of a reference mechanical state, linked to osteocyte-specific biochemical and morphological parameters; (ii) it accounts for a rapid decrease in mechanical sensitivity, linked to the partial desensitisation of osteocytes; (iii) it accounts for a slow resetting of a mechanical setpoint memory by bone remodelling, linked to the replacement of osteocytes with ones adapted to current mechanical states. Both short-term and long-term adaptation responses are thus accounted for and explicitly associated with distinct physiological processes.

We find that bone's long-term mechanical memory in a \rve\ accommodates to new mechanical environments at a rate entirely determined by bone formation rate and bone volume fraction, see Eq.~\eqref{SEDcell_ODE}. This equation is found by averaging microscopic evolution equations of remodelling. It has no new independent parameters and as such, cannot be calibrated. This is a significant difference compared to previous phenomenological models that have interpreted a long-term relaxation of the setpoint by cell accommodation and have introduced unrealistically slow accommodation rate parameters in a similar equation~\citep{Garcia-Aznar2005, Schriefer2005}.

The space-dependent and time-dependent mechanical reference state that we define helps to alleviate important shortcomings of Frost's mechanostat theory. The mechanical memory recorded in the osteocyte network leads to hysteresis, i.e., a loading-history dependence of bone structures. Near the neutral axis of long bones, small values of mechanical stimulus prevailing in these regions gradually become recorded as a new reference mechanical state during tissue renewal, preventing total loss of bone in this region. The fact that osteocyte replacement only occurs during bone formation could be an important factor in explaining asymmetries of bone responses during mechanical regulation. These asymmetries depend on how fast pre-existing populations of osteocytes are replaced by the remodelling process, which in turn depends on turnover rate, bone microstructure, and thus bone anatomical site. Asymmetries between bone gain and bone loss responses in our model also depend on the strength of mechanotransduction encoded in the parameters $\beta_\ob$, and $\beta_\oc$ and, importantly, on the geometrical regulation of bone cell activation by the microstructure, which plays a dominant role~\citep{Martin1972,Buenzli2013,Lerebours2015}.

The ability of our osteocyte-based mechanostat to capture dynamic responses at different timescales is exemplified by the simulations of long-term disuse and exercise regimens. In long-term disuse, the cell replacement mechanism enables bone structure to stabilise within 10 years after 30\% bone loss (Figure~\ref{Disuse30y_comparison_models}). The final value of $\fbm$ depends on the initial value of \fbm, which represents bone type specificity, and on the loading history. This is in qualitative and quantitative agreement with experimental data~\citep{Schriefer2005, Vico2000, Dudley-Javoroski2012, Edwards2015a}. In contrast, without replacement of the long-term mechanical setpoint by remodelling, stabilisation occurs much later and the total loss of bone is always proportional to the decrease in applied loads. Indeed, if stabilisation is only driven by changes in microstructure, and not by changes in the mechanical memory encoded in osteocytes as a bone material property, then the final state is reached when $\Psi$ in disuse returns to its initial value $\Psi(t\!=\!0)$. By Eq.~\eqref{Psi_fct_fbm}, this occurs only once the ratio $F_\text{disuse}/\fbm$ returns to its initial value $F/\fbm(t\!=\!0)$, i.e., once $\fbm/\fbm(t\!=\!0)=F_\text{disuse}/F$.

The frequency and timing of small periodic loadings simulating exercises has a marked effect on bone gain (Figure~\ref{Influence_periods}). Bone gain is enhanced by rest periods, because it allows osteocytes to re-sensitise and makes them respond more strongly to a new overload. Our results agree qualitatively with studies that have shown increased gain by exercising more often for shorter periods~\citep{Ehrlich2002, Srinivasan2007, Thompson2012}. Nonstop exercise in our model still leads to bone gain, as seen in animal studies of hypergravity~\citep{Gnyubkin2015}. However, some studies have found that exercising more than 8 hours daily does not increase bone~\citep{Burr2002, Robling2001, Turner2004}. Cells at that point are mostly desensitised. Bone gain is driven by the much slower long-term dynamics of structural and material adaptation, and may have been missed by these studies. 

The asymmetry in the transduction response $\mu(t)$ between overload and return to normal load in Figure~\ref{Bumping10d} emphasises the distinction between a desensitised state in overload and a desensitised state under normal load. This distinction reflects the long-term memory of the reference mechanical state encoded in the osteocytes, and is responsible for the gain of bone. When jumping from overload to normal load, the short-term transduction response $\mu(t)$ becomes negative (Figure~\ref{Bumping10d}). It is qualitatively similar to an underload response. Experimentally, both re-sensitisation periods and underload induce the adaptation of links between cells and their extracellular matrix~\citep{Rubin1999}.

Daily exercises superimposed to long-term disuse (Figure~\ref{Disuse30y_comparison_models}) are able to curb bone loss significantly in the model. Bone is lost more slowly if exercises are started immediately post injury, in agreement with clinical data~\citep{Astorino2013, Dudley-Javoroski2012}. In our model, recovery is almost as complete if exercises are started only after 10 years, i.e., once the bone contains new osteocytes adapted to the reduced mechanical environment. Exercising immediately post injury was suggested to prevent the loss of key structural elements~\citep{Coupaud2009, Edwards2014, Edwards2015a}. This cannot be investigated by our current model, which only considers bone volume fraction in a \rve\ of the tissue.

In conclusion, the explicit consideration of the osteocytic basis of mechanosensation and mechanotranduction enables us to interpret different timescales of bone adaptation by the physiological processes of osteocyte desensitisation and osteocyte replacement during bone remodelling. While the osteocyte-based mechanostat theory we propose is constructed from experimental facts, it requires more direct experimental verification. The mathematical model we used to test this theory relies on a number of assumptions. Osteocytes are assumed to be formed adapted to their mechanical environment instantly, but they are known to mature in multiple stages. Some elements encoding the reference mechanical state could be set early (e.g. the morphology of the lacuno-canalicular network), while others could be set later (e.g. gene expression at osteocyte maturity). A number of physiological processes were not considered explicitly in the model. Osteocytic osteolysis could result in changes in the long-term mechanical setpoint but it was not considered because it is believed to be driven by calcium homeostasis rather than mechanical feedback, and it is only reported to result in small changes in volumes (expansion/contraction in lacuna volume and canaliculi diameters)~\citep{Atkins2012,Vatsa2008, vanHove2009,Han2004}. Osteocytes have been reported to dynamically expand their dendrites within the canaliculi and create and remove connexions in neonatal mouse calvaria~\citep{Dallas2010}. This may provide a further mechanism to modulate the setpoint on fast timescales. We did not consider this possibility explicitly in our model as its link with mechanical sensitivity is unclear, and it is yet unproven that this behaviour holds in mature osteocytes and in osteocytes in adult bone. Finally, our model did not consider changes in osteocyte density in new bone, nor osteocyte apoptosis, which is known to increase with age. More extensive testing of the proposed mechanostat in a spatio-temporal framework will be the subject of future work.

\section*{Acknowledgements}
We thank Natalie~A~Sims and Mark~R~Forwood for helpful and stimulating discussions. PRB gratefully acknowledges the Australian Research Council for Discovery Early Career Researcher Fellowship (project number DE130101191).

\section*{Conflict of interest statement} None

\appendix
\section{Model description}\label{sec:model-description}
This Appendix details the mathematical model summarised in Section~\ref{Sec_math_model}. The cell population model that describes the evolution of active osteoblasts and active osteoclasts under biochemical, geometrical, and mechanical regulations in a region of interest (\rve) of the tissue is presented in \ref{subsection_bone_remodelling}. The cellular scale model of mechanobiological transduction and cell desensitisation, described by receptor--ligand signalling and trafficking within an osteocyte, is presented in \ref{subsection_cell_desensit}. Osteocyte replacement is described by averaging microscopic governing equations of tissue modelling and remodelling in the \rve\ in \ref{subsection_cell_replacement}. The calculation of mechanical stimulus from applied loads and bone volume fraction is presented in \ref{subsection_meca}. Finally, \ref{subsection_Calib} details aspects related to numerical simulations and model calibration.

\subsection{Generic bone cell population model}\label{subsection_bone_remodelling}
Bone volume fraction is an important microstructural characterisation of the \rve\ influencing the local mechanical state~\citep{Grimal2011, Granke2012, Rohrbach2012}. The balance of $\fbm$ due to bone formation and resorption is given by
\begin{align}\label{ODE_fbm}
\frac{\partial}{\partial t} \fbm (t) = \kform ~\ob - \kres ~\oc,
\end{align}
where $\ob$ and $\oc$ are the number of active osteoblasts and active osteoclasts per unit volume in the \rve, and $\kform$ and $\kres$ are the volume of bone formed and resorbed per cell per unit time, respectively~\citep{Martin1972,Pivonka2008,Buenzli2013,Lerebours2015}. Biochemical, geometrical, and mechanical regulations are assumed to influence the cell populations as follows:
\begin{align}
 \kform\,\ob &=\sv(\fbm)(1\!-\!\fbm) [\alpha_{\ob} + \beta_{\ob}\,  \fbm\, \chi_\ob(\mu)], \label{kfOB}\\
 \kres\,\oc &=\sv(\fbm)(1\!-\! \fbm) [\alpha_{\oc} + \beta_{\oc}\, \fbm\,\chi_\oc(\mu)]\label{krOC}.
\end{align}
The density of bone surface $\sv$ quantifies the propensity of the \rve\ to undergo remodelling\footnote{In standard bone histomorphometric notations, $\fbm=\bv/\tv$ and $\sv=\bs/\tv$~\citep{Dempster2013}.}~\citep{Martin1972,Pivonka2013,Buenzli2013, Lerebours2014}. The factor $\sv(\fbm)(1\!-\!\fbm)$ represents the geometrical influence of the \rve's microstructure on the activation of osteoblasts and osteoclasts. It corresponds to the probability that precursor cells living in the pore volume (factor $1-\fbm$) generate active cells at the bone surface (factor $\sv$). We use the relationship $\sv(\fbm)$ proposed by Martin~\citep{Martin1984, Buenzli2013}. The proportionality coefficients $\alpha_\ob,\alpha_\oc>0$ represent the biochemical influence on the activation rates of osteoblasts and osteoclasts. The $\mu$-dependent terms represent the mechanical regulation of these activation rates. The parameters $\beta_\ob$ and $\beta_\oc$ modulate the strength of mechanical regulation in formation and resorption. Because osteocytes form a densely interconnected network likely to integrate mechanical stimulus over large spatial scales~\citep{Marotti2000, Adachi2010, Buenzli2015,Buenzli2015a}, mechanical regulation is assumed to be proportional to the population of osteocytes in the $\rve$ (factor $\fbm$) and to the biochemical stimulus $\mu$ (see functions $\chi_\ob, \chi_\oc$ below). Experimentally, in overload, the population of osteoblasts increases without significant change in osteoclasts, and in underload, the population of osteoclasts increases without significant change in osteoblasts~\citep{Chow1998, Burger1999, Huiskes2000, Turner2004}. Accordingly, we set
\begin{align}\label{mech-reg}
    \chi_\ob(\mu) = \begin{cases}0, &\ \mu < 0,\\\mu, &\ \mu >0.\end{cases} 
\quad 
\chi_\oc(\mu) = \begin{cases} |\mu|, &\ \mu < 0,\\0, &\ \mu > 0.\end{cases}
\end{align}

In a healthy state, biochemical and mechanical regulations are in equilibrium, i.e., $\alpha_{\ob} = \alpha_{\oc} \equiv \alpha$ and $\mu=0$. In this case, the \rve's bone volume fraction $\fbm$ in Eq.~\eqref{ODE_fbm} is constant, but continually turned over with rate
\begin{equation}\label{turnover}
\alpha\, (1 \!-\! \fbm)\,\sv(\fbm)
\end{equation}
(in volume fraction per unit time)~\citep{Lerebours2015}.

\subsection{Osteocyte model of mechanobiological transduction and desensitisation}\label{subsection_cell_desensit}
The long-term mechanical setpoint $\SEDcell$ introduced in Section \ref{section_proposal_mechanostat} is defined as the tissue average of long-term mechanical memories $\SEDcellmicro$ encoded in individual osteocytes:
\begin{equation}
\SEDcell(t) = \langle \SEDcellmicro \rangle_{\bv} = \frac{1}{\bv(t)} \int_{\bv(t)}\hspace{-1.5em} \textrm{d}^3r ~\SEDcellmicro(\b r,t), \label{SEDcell_def}
\end{equation}
where $\bv(t)$ is the bone volume in a fixed \rve\ of the tissue. We propose below a cellular scale model by which an individual osteocyte senses and transduces mechanical stimulus, and partially accommodates to it. This model shows in particular that morphological and genotypic properties of an osteocyte can provide a material basis for $\SEDcellmicro$. The modulation of these properties during the osteocyte's formation is thus a mechanism by which long-lasting mechanical memories can be recorded in bone.

We assume that an osteocyte first transduces the mechanical stimulus $\Psi$ into an intracellular agent $S$ (Figure~\ref{cell_desensitisation_model}). If $S$ is greater than a reference level $S_0$, the osteocyte considers itself overloaded. If $S$ is lower than $S_0$, the osteocyte considers itself underloaded. The reference level $S_0$ is assumed to be a long-lasting genotypic property. It will be related below to the osteocyte's long-term setpoint \SEDcellmicro. We assume that intracellular signalling cascades are triggered such that (i) in overload, the osteocyte generates an extracellular biochemical stimulus $\mu >0$; and (ii) in underload, the osteocyte generates $\mu <0$. This extracellular biochemical stimulus is defined in our model by:
\begin{align}
    \mu = m(S-S_0).\label{mu}
\end{align}
This stimulus is assumed to represent the mechanical signal received by active osteoblasts and osteoclasts. This signal is also assumed to feed back into the transduction mechanism to allow partial desensitisation of the cell to $\Psi$. We consider a single transduced biochemical stimulus for simplicity. More realistically, mechanobiological transduction could involve distinct signalling pathways in overload, underload, and cell accommodation. 

The transduction from $\Psi$ to $S$ is assumed to occur through mechano-receptors $R$ (Fig~\ref{cell_desensitisation_drawing}). Following the stimulation of $R$ by $\Psi$, cellular mechanisms are assumed to generate compounds $S$, and to produce or degrade components $R$ in response to $S$ (Fig~\ref{cell_desensitisation_equation}). The dynamic nature of the number of $R$s enables the cell to adapt its sensitivity to $\Psi$~\citep{lauffenburger-lindenman}. This model of mechanobiological transduction and desensitisation is formulated as a system of receptor--ligand reactions, governed by:
\begin{align}
\frac{\partial S}{\partial t} &= \f(\Psi) \, R - D_{S} \, S \label{dsdt},\\
\frac{\partial R}{\partial t} &= - ~A\, \mu - D_{R}\,  R + P_{R} \label{dRdt}.
\end{align}
The first term in the right hand side of Eq.~\eqref{dsdt} describes the transduction of $\Psi$ to $S$ through the mechano-receptors $R$. It involves an increasing function $f(\Psi)$ specified in more detail below. The second term is a first-order degradation rate that prevents unbounded accumulation of $S$. The first term in the right hand side of Eq.~\eqref{dRdt} describes the desensitisation of the cell to $\Psi$: if $\mu>0$, mechano-receptors $R$ are removed, which decreases the transduction of $\Psi$ to $S$ in Eq.~\eqref{dsdt} and lowers $\mu$; if $\mu<0$, mechano-receptors $R$ are produced, which increases the transduction of $\Psi$ to $S$ and raises $\mu$. The subsequent terms in Eq.~\eqref{dRdt} correspond to a degradation and production of $R$ to ensure baseline amounts of mechano-receptors. These terms are required to prevent the total accommodation of the cell's mechano-receptors to $\Psi$~\citep{lauffenburger-lindenman}. After rapid transients, the number of mechano-receptors stabilises and the cell is partially desensitised to $\Psi$. The biochemical stimulus $\mu$ relaxes to the steady-state value
\begin{align}
\overline{\mu} &=  m\frac{P_{R}\, \f(\Psi) - D_S\, D_R \,S_{0}}{m \,A\,\f(\Psi) + D_S \,D_R}, \label{mu_bar}
\end{align}
found from Eqs~\eqref{dsdt}--\eqref{dRdt}. This steady-state value retains a dependence on the mechanical stimulus $\Psi$.

We note that Equations~\eqref{dsdt}--\eqref{dRdt} could also be interpreted as a model of desensitisation of the mechano-receptors themselves. Removal and creation of receptors is mathematically equivalent to switching receptors to inactive and active states, respectively. In either case, the cell's mechanical sensitivity is modified. Several concurrent biological mechanisms of osteocyte desensitisation may actually be involved (see Section \ref{subsection2.1_cell_desensit}). For simplicity, only one is considered in this mathematical model.

\subsubsection*{Cellular long-term reference mechanical state $\textup{\SEDcellmicro}$}
The long-term value $\overline{\mu}$ of $\mu$ has a sign that depends on the function $\f$ and on the cell-specific parameters $P_R$, $D_S$, $D_R$, and $S_0$ occurring in the numerator of Eq. \eqref{mu_bar}. These model elements enable us to define a long-term cellular reference mechanical state $\SEDcellmicro=\SEDcellmicro(\f, P_R, D_S, D_R, S_0)$ by inverting
\begin{align}\label{sedcell-def}
    \f(\SEDcellmicro) = \frac{D_S D_R S_0}{P_R},
\end{align}
such that if $\Psi>\SEDcellmicro$ the osteocyte is overloaded and if $\Psi<\SEDcellmicro$ the osteocyte is underloaded. With the definition \eqref{sedcell-def}, one has
\begin{align}\label{proxy}
\Psi \gtreqqless \SEDcellmicro \iff \overline\mu \gtreqqless 0.
\end{align}
Equations~\eqref{mu}--\eqref{dRdt} can be recast into the differential equation of a forced damped oscillator $\mu(t)$ with $\Psi$-dependent frequency, damping, and forcing. To prevent transient oscillations, we assume an overdamped regime, which imposes the constraint $  2\sqrt{mA\f(\Psi)} \leq D_S - D_R$ at all times. We thus assume $\f$ to be bounded, and given by:
\begin{align}\label{f}
    \f(\Psi) = \f_0 \frac{K+k_\Psi \Psi}{1+k_\Psi \Psi},
\end{align}
where $K<1$ for $\f$ to be an increasing function of $\Psi$, and $K>0$ to ensure that $\f(0)\neq 0$, which is needed for the equivalence~\eqref{proxy}: it prevents the neutral axis (where $\Psi=0$) from being always resorbed irrespective of $\SEDcellmicro$, see Eq.~\eqref{mu_bar}.

\subsection{Osteocyte replacement and renewal of~$\textup{\SEDcellmicro}$}\label{subsection_cell_replacement}
The mechanostat's long-term setpoint $\SEDcellmicro$ is defined by osteocyte-specific morphological and genotypic parameters (protein production and degration rates) via Eq.~\eqref{sedcell-def}. The function $\f$ is associated with the efficiency with which mechanical stimulus $\Psi$ is sensed by the osteocyte. This efficiency likely depends on the morphology of the lacunar cavity containing the osteocyte. Based on the idea that this will not change significantly until the tissue is removed and replaced, we choose to record the value of $\SEDcellmicro$ into the function $f$ through the parameters $k_\Psi$ and $K$ by setting
\begin{align}
k_\Psi(\SEDcellmicro) &= \frac{\frac{D_SD_RS_0}{P_R f_0} - K(\SEDcellmicro) }{\SEDcellmicro\left(1-\frac{D_SD_RS_0}{P_R f_0}\right)}, \label{def_kpsi}\\
\quad K(\SEDcellmicro) &= \frac{D_SD_RS_0}{P_R f_0} \frac{1}{1+\gamma_K \SEDcellmicro} \label{def_K}
\end{align}
such that Eq.~\eqref{sedcell-def} holds and the condition $0<K<1$ is fulfilled. 

These parameters are assumed to be set during the osteocyte's formation and to remain unchanged until the osteocyte's removal. Since bone undergoes periodic remodelling, these parameters' values can be updated. This provides a mechanism to modulate the setpoint over much slower timescales than the cellular adaptation of mechano-receptors that occurs during the osteocytes' lifetime. We thus view $\SEDcellmicro$ in Eqs \eqref{def_kpsi}--\eqref{def_K} as a long-lasting bone tissue property and estimate the evolution of its mean value in the \rve\ ($\SEDcell$, defined by Eq.~\eqref{SEDcell_def}), during the renewal of bone matrix, which replaces osteocytes. For this, we average spatio-temporal equations governing the evolution of tissue properties at the cellular scale under bone modelling and remodelling~\citep{Buenzli2015b}. 

At the cellular scale, the evolution of \SEDcellmicro\ due to formation and resorption processes is governed by:
\begin{equation}
\frac{\partial}{\partial t} \SEDcellmicro(\b r, t) = \Psi ~v_{\textrm{form}} ~\updelta_{S(t)} - \SEDcellmicro ~v_{\textrm{res}} ~\updelta_{S(t)}, \label{ODE_SEDcellmicro}
\end{equation}
where $S(t)$ is the bone surface in the \rve, $\updelta_{S(t)}$ is a surface distribution (formally zero everywhere except at $S(t)$ where it is infinite), and $v_{\textrm{form}}=\kform \rho_\ob$ and $v_{\textrm{res}}=\kres \rho_\oc$ are the normal velocity of $S(t)$ on formation and resorption surfaces, which depend on the surface densities of osteoblasts and osteoclasts $\rho_\ob$ and $\rho_\oc$~\citep{Buenzli2015b}. The first term in the right hand side of Eq.~\eqref{ODE_SEDcellmicro} records the current value of the mechanical stimulus $\Psi$ into the value $\SEDcellmicro(\b r, t)$ set through modulation of $k_\Psi$ and $K$ during new bone formation at $\b r$. Because $\SEDcellmicro$ vanishes out of the bone volume $\bv(t)$, the integral in Eq.~\eqref{SEDcell_def} can be extended to the constant \rve\ integration domain $\tv$ (tissue volume~\cite{Dempster2013}). Differentiating with respect to $t$ provides an evolution equation for the mean value $\SEDcell$ which cannot be written in closed form because resorption proceeds from the bone surface and removes values of $\SEDcellmicro$ recorded last, rather than the current average $\SEDcell$~\citep{Buenzli2015b}. Provided that the distribution of $\SEDcellmicro$ in the tissue is not too heterogeneous, the mean-field approximation
\begin{align}\label{m-f}
    \SEDcellmicro \approx \langle \SEDcellmicro \rangle_{\bv} = \SEDcell
\end{align}
is well satisfied and enables closure of the evolution equation for $\SEDcell$. This approximation is expected to hold for large-enough \rve s containing several concurrent resorption events removing values close to $\SEDcell$ in average.
Using $\bv(t) = \fbm(t) \tv$, Eq.~\eqref{ODE_fbm}, Eq.~\eqref{ODE_SEDcellmicro}, the mean-field approximation~\eqref{m-f}, and resolving volume integrals over the surface distribution as surface integrals gives Eq.~\eqref{SEDcell_ODE} (see Ref.~\cite{Buenzli2015b} for more details):
\begin{align}
    \pd{}{t}\SEDcell(t) = \frac{1}{\fbm} \kform \langle \ob \rangle_\tv \left(\Psi - \SEDcell\right).
\end{align}

\subsection{Determination of the mechanical stimulus}\label{subsection_meca}
The mechanical stimulus $\Psi$ sensed by osteocytes is assumed in this paper to be the microscopic strain energy density of the bone matrix phase in the \rve, defined by
\begin{align}
\Psi &= \frac{1}{2} \b \varepsilon^{\micro}_\bm : \cmicro_\bm : \b \varepsilon^{\micro}_\bm \label{def_SEDmicro},
\end{align}
where $\cmicro_\bm$ is the bone matrix stiffness tensor, and $\b \varepsilon^{\micro}_\bm$ the microscopic strain of bone matrix~\citep{Lerebours2015}. Because the bone in the \rve\ is porous, microscopic stresses of bone matrix are larger than the tissue-average stress $\langle\sigma^\text{micro}\rangle_\tv = F/L^2$ induced by the compressive force $F$ onto the \rve\ ($L^2$ is the \rve's cross-sectional area). This stress concentration effect can be estimated by micromechanical homogenisation schemes~\citep{Zaoui2002, Hellmich2008}. A simpler approach is to assume that for small deformation, microscopic strains of the bone matrix phase and of the vascular (soft tissue) phase in the \rve\ coincide with tissue-average strains:
\begin{equation}\label{approx_epsilon}
\varepsilon^\micro_\bm \approx \varepsilon^\micro_\vas \approx \langle\varepsilon^\text{micro}\rangle_\tv.
\end{equation}
With the assumption that the mechanical state of each phase in the \rve\ is homogeneous, $\langle\sigma^\text{micro}\rangle_\tv = \fbm \sigma^\text{micro}_\bm + \fvas \sigma^\text{micro}_\vas$. Using Eqs~\eqref{def_SEDmicro},\eqref{approx_epsilon} with Hooke's law and taking $\cmicro_\vas = 0$ so that the vascular phase does not bear loads, gives the following explicit dependence of $\Psi$ upon $\fbm$:
\begin{equation}\label{Psi_fct_fbm}
\Psi = \frac{1}{2} {\cmicro_\bm}^{-1} \Big(\frac{F}{L^2 \fbm}\Big)^2.
\end{equation}
A comparison of this formula with numerical evaluations of a micromechanical homogenisation scheme in the case of pure compression showed a maximum discrepancy of 3\% when $\fbm\to 0$. This discrepancy was attributable to difficult numerical evaluations of the micromechanical formulas at such low $\fbm$. We note that in this regime, micromechanical formulas are only formal as the theory may not be valid.

\subsection{Numerical simulations and model calibration}\label{subsection_Calib}
\begin{table*}[t]
\begin{center}
\caption{Model parameters}{\small
\begin{tabular}{@{}l@{}l@{}p{1.3\columnwidth}@{}}
\toprule
Symbol& \hspace{2em} Value & \hspace{2em} Description\\
\thickmidrule
\bstrut{1.5ex}$\alpha$ & \num{2e-4}\,mm/day & Biochemical regulation  for osteoclasts and osteoblasts (calibrated)
\\ $\beta_{\oc}$ & 0.175\,mm/day/cell & Strength of the mechanical regulation  for osteoclasts (calibrated)
\\ $\beta_{\ob}$ & 1\,mm/day/cell & Strength of the mechanical regulation  for osteoblasts (calibrated)
\\\midrule
$A$ & \num{1.23e5}\,$\textrm{day}^{-1}$ & Cell desensitisation rate (calibrated)
\\ $m$ & \num{3e-4}\, & Biochemical transduction parameter (parametric study)
\\ $D_S$ & 100\,$\textrm{day}^{-1}$ & Degradation rate of $S$ (parametric study)
\\ $D_R$ & 2\,$\textrm{day}^{-1}$ & Degradation rate of $R$ (parametric study)
\\ $P_R$ & 1000\, $\textrm{cell}^{-1}\textrm{day}^{-1}$ & Production rate of $R$ (arbitrary scaling parameter)
\\ $\gamma_K$ & $34 10^4$ $\textrm{GPa}^{-1}$ & Parameter of $K(\SEDcellmicro)$ (parametric study)
\\ $f_0$ & \num{20.5}\,$\textrm{day}^{-1}$ & Mechanical transduction parameter (parametric study)
\\ $S_{0}$ & 100\,$\textrm{cell}^{-1}$ & Equilibrium number of $S$ (arbitrary scaling parameter)
\\\midrule
$F/L^2$ & 30\,MPa & Applied macroscopic stress \citep{Pivonka2013, Scheiner2013} ($F = 120$N and $L^2 =$4\,$\textrm{mm}^{2}$ \cite{Grimal2011, Lerebours2014}).
\\ $\cmicro_\bm$ & 28.4 GPa & Bone matrix stiffness \citep{Ashman1984, Fritsch2007}
\\\bottomrule
\end{tabular}}
\label{Parameter_table}
\end{center}
\end{table*}

Equations~\eqref{ODE_fbm}--\eqref{mech-reg},\eqref{mu}--\eqref{dRdt}, and \eqref{SEDcell_ODE} govern the evolution of the \rve's bone volume fraction $\fbm$ for a given mechanical stimulus $\Psi$. In turn, $\Psi$ is determined by $\fbm$ and the external force applied on bone, Eq. \eqref{Psi_fct_fbm}, which closes the system of equations. These equations were integrated numerically using \texttt{Matlab}'s ODE solver \texttt{ode15s} with the parameters listed in Table~\ref{Parameter_table}.

Our mathematical model contains 11 parameters to describe the regulation of bone under biochemical, geometrical, and mechanical regulations, and two parameters to describe external macroscopic stress and bone matrix stiffness. The latter two parameters were taken from the published literature (see Table~\ref{Parameter_table}).

The parameters $\alpha$, $\beta_\ob$, and $\beta_\oc$ were calibrated as follows. The value of $\alpha$ in Eq.~\eqref{turnover} determines the rate of bone turnover in health. It also influences the long-term relaxation rate of the mechanical reference state to new loads in mechanical disuse and overuse, induced by osteocyte replacement in Eq.~\eqref{SEDcell_ODE}. We therefore calibrated $\alpha$ to obtain rates of turnover compatible with Parfitt's measurements~\citep{Parfitt1983,Lerebours2015} and such that the mechanical reference state $\SEDcell$ reaches a new accommodated value within 10\,years in simulations of mechanical disuse. The parameters $\beta_\oc$ and $\beta_\ob$ determine the strength of mechanical regulation. They were calibrated such that bone is lost at a rate of 0.3\% per month during long-term spaceflight missions, with no significant gain 6 months after the return to Earth~\citep{Collet1997, Vico2000}.

A scaling analysis of the cell desensitisation model was performed, showing that the number of compounds $S$ and the number of mechano-receptors $R$ in the osteocyte can be scaled arbitrarily without modifying the model's behaviour~\citep{Barenblatt1996}. Time was not scaled to retain its physiological meaning, but the cell desensitisation rate $A$ was calibrated such that $\mu(t)$ relaxes to $\overline\mu$ in 8\,hours upon an increase in $\Psi$ when simulating exercises starting from a healthy state~\citep{Robling2001, Burr2002, Turner2004}. The arbitrary scaling on $S$ and $R$ were set such that in mechanical equilibrium, the number of compounds $S$ in the osteocyte is $\overline{S}=S_0=100/\text{cell}$, and the number of mechano-receptors $R$ is $\overline{R}=P_R/D_R=500/\text{cell}$. The parameters $S_0$ and $P_R$ were thereby chosen to be arbitrary scaling parameters. The remaining five parameters $D_R, D_S, m, f_0$, and $\gamma_K$ were determined through a parametric study subjected to the constraints $(D_S - D_R)^2\geq 4\,m\,A\,f_0$ to prevent oscillatory behaviour, and $\frac{D_SD_RS_0}{P_R f_0} < 1$ to ensure that $K(\SEDcellmicro)<1$. The value $D_S D_R S_0/P_R$ in the right hand side of Eq.~\eqref{sedcell-def} was chosen to be $0.97 f_0$, allowing for sufficient mechanical sensitivity of the transduction. Due to the calibration of $\alpha$, $\beta_\ob$, and $\beta_\oc$ the choice of these remaining 5 parameters had little impact on the model's behaviour at the tissue scale.

Mechanical disuse and overuse were simulated by decreasing and increasing the external applied force $F$. Mechanical disuse simulations considered a reduction in $F$ to one third of its initial value. This reduction is estimated to correspond to low gravity settings during long-term spaceflight missions and to reduced loads in legs of spinal cord injury patients. Simulations only depends weakly on this assumption due to the calibration of $\beta_\ob$ and $\beta_\oc$. Mechanical overuse during exercise simulations considered an increase in $F$ of 20\%.

\section*{References}
\bibliographystyle{abbrvnat} 
\bibliography{My-Collection}

\end{document}